# Average amplitude and phase detuning near driven 3rd integer resonance

Philipp Niedermayer

## 1 Introduction

We consider the particle dynamics near a 3rd order resonance driven by sextupole magnets, such as commonly used for resonant slow extraction of accelerated particle beams from synchrotrons [Ben99]. The dynamics under these conditions are dominated by non-linear particle motion and lead to detuning effects, typically described as detuning with amplitude [Fei94; Whi13].
However, the detuning depends not only on the particle amplitude, but also on the angle in phase space. This is not only a consequence of the triangularly deformed phase space trajectories of the particles, where the amplitude (or action) itself depends on the angle. Moreover, the effect of the sextupole kick is – depending on the angle – either aligned or in opposite direction with the transverse momentum, causing the phase advance for the subsequent turn to be in- or decreased.

In this note, we derive an expression for the average phase advance of a particle, the trajectory of which is governed by the Kobayashi Hamiltonian. The Hamiltonian is a measure for the oscillation energy and – in contrast to the amplitude – a quantity conserved in time.

### 1.1 Kobayashi theory

The Kobayashi theory [Kob70] describes the particle dynamics in the close vicinity of a driven 3rd order resonance. One defines the (linear) fractional tune as $q = r + d$, where $r = 1/3$ or $2/3$ is the resonance and the distance $d \ll 1$ is small. The resonance is driven by sextupole magnets, whose combined effect is summarized by a virtual sextupole magnet with normalized strength $S = -\frac{1}{2}\beta_x^{3/2}k_2 l$ in $\mathrm{m}^{-1/2}$ [Ben97, pp. 30 sqq.].
It is convenient to use normalized horizontal phase space coordinates $X$ and $X'$ (in units of $\mathrm{m}^{1/2}$): [Wie15, pp. 115, 233 sqq.]:

$$\begin{pmatrix} X \\ X' \end{pmatrix} = \begin{pmatrix} 1/\sqrt{\beta_x} & 0 \\ \alpha_x/\sqrt{\beta_x} & \sqrt{\beta_x} \end{pmatrix} \begin{pmatrix} x \\ x' \end{pmatrix} \tag{1}$$

where $\alpha_x$ and $\beta_x$ are the twiss parameters, $x$ is the horizontal displacement in m and $x'$ the divergence. Equivalently, the phase space can be parametrized in polar coordinates by means of the action $J$ and angle $\theta$:

$$X = \sqrt{2J}\cos(\theta) \qquad\qquad X' = -\sqrt{2J}\sin(\theta) \tag{2}$$

The Kobayashi Hamiltonian $H$ [Kob67, p. 348] describes the three turn motion of a particle at the location of the sextupole:

$$H = 3\pi d\left(X^2 + X'^2\right) + \frac{S}{4}\left(3XX'^2 - X^3\right) \tag{3}$$

$$= 6\pi d J - \frac{S}{\sqrt{2}} J^{3/2}\cos(3\theta) \tag{4}$$

The phase space trajectories (equipotential lines) have a characteristic triangular shape, as depicted in figure 1 (left). The separatrix at $H_{\text{sep}} = (4\pi d)^3/S^2$ marks the border of stable motion. The size of this stable equilateral triangle is given by the radius of the inscribed circle $h = 4\pi d/S$.

## 1.2 Detuning

According to Hamilton's equation, the three turn phase advance, i.e. the change of angle $\theta$ per three turns, is given as

$$\mu_3 = \frac{\partial H}{\partial J} = 6\pi d - \frac{3S}{4}\sqrt{2J}\cos(3\theta) = \frac{3H}{2J} - 3\pi d \tag{5}$$

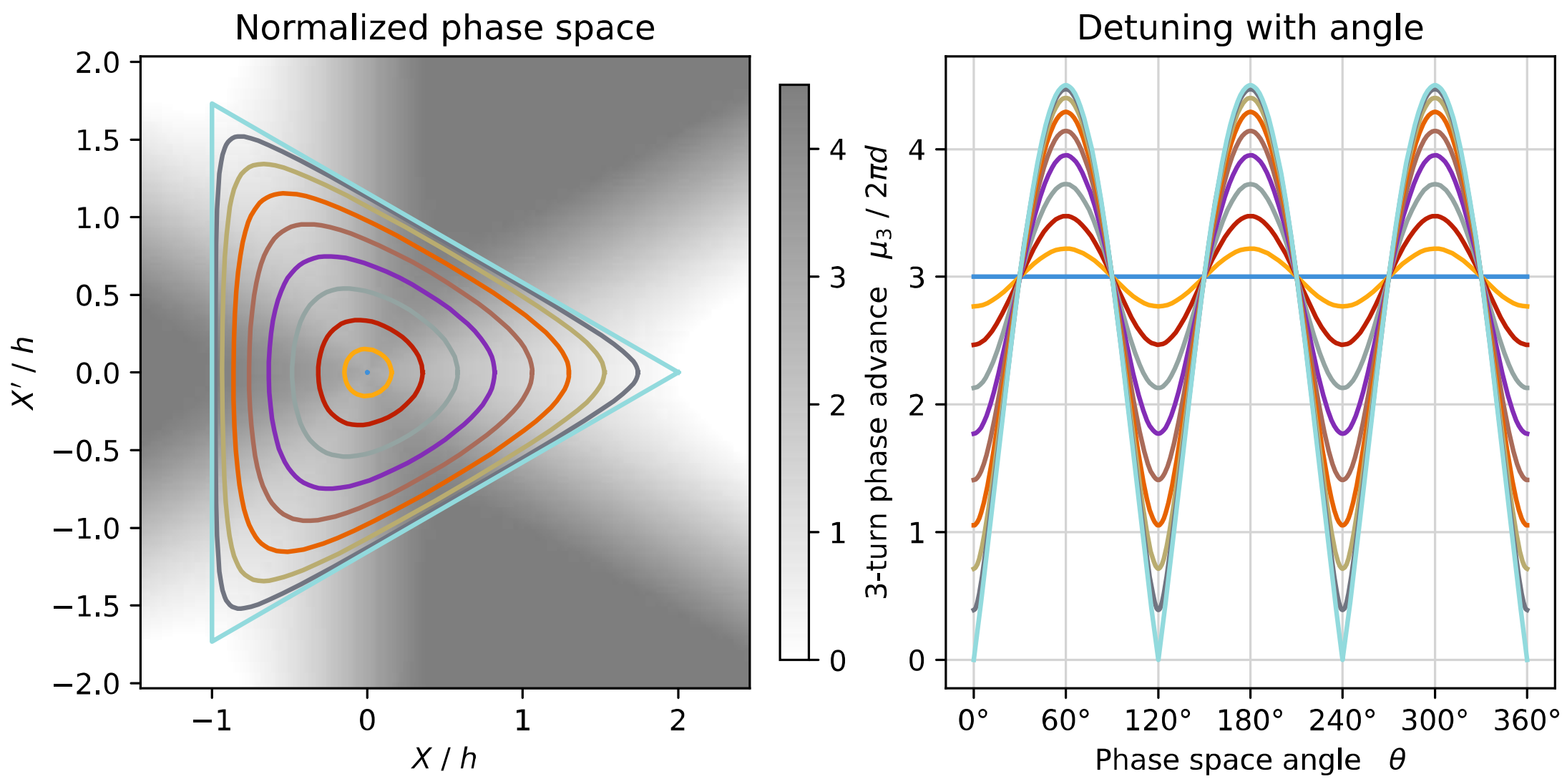

Figure 1: Left: Normalized phase space with equipotential lines of the Kobayashi Hamiltonian. The colour shading indicates the detuned three turn phase advance. Right: three turn phase advance as function of angle for identical values of $H$.

The term $6\pi d$ is the three turn phase advance (modulo $2\pi$) for a linear system ($S = 0$). The additional term describes the nonlinear detuning which depends not only on the amplitude $\sqrt{2J}$, but also on the angle $\theta$ in phase space [Cor22]. The detuning with angle is depicted in figure 1 (right).

# 2 Average detuning

As the particle moves along the equipotential lines of the Hamiltonian over many turns, the three turn phase advance varies significantly. In the following, we derive an expression for the average phase advance a particle with given $H$ experiences over many turns. Therefore, we will consider a number of turns $n \gg 1/d$, such that the particle has sampled the full angle along the equipotential line.

## 2.1 Time average over many turns

The time average of a quantity $g$ is given as:

$$\langle g \rangle = \frac{\int g \, \mathrm{d}t}{\int \mathrm{d}t} = \frac{\int g \, \mathrm{d}n}{\int \mathrm{d}n} \tag{6}$$

where the integral over time $t$ was written as an integral over the turn number $n = f_{\mathrm{rev}} t$ with the revolution frequency $f_{\mathrm{rev}}$. Using the three turn phase advance $\mu_3 = \mathrm{d}\theta \,/ (3 \, \mathrm{d}n)$, the expression can be transformed into an integral over the phase space angle:

$$\langle g \rangle = \frac{\int \frac{g}{\mu_3} \, \mathrm{d}\theta}{\int \frac{1}{\mu_3} \, \mathrm{d}\theta} \tag{7}$$

This is valid, because near the 3rd order resonance the change in angle per 3 turns is in the order of $6\pi d \ll 1$ and can therefore be considered infinitesimal.
With $g = \mu_3$ and the integration over the full angle as motivated earlier, one gets:

$$\langle \mu_3 \rangle = \frac{\oint \mathrm{d}\theta}{\oint \frac{1}{\mu_3} \, \mathrm{d}\theta} = \frac{2\pi}{\oint \frac{1}{\mu_3} \, \mathrm{d}\theta} \tag{8}$$

or for an arbitrary quantity $g$:

$$\langle g \rangle = \frac{\langle \mu_3 \rangle}{2\pi} \oint \frac{g}{\mu_3} \, \mathrm{d}\theta \tag{9}$$

## 2.2 Average phase advance as function of the Hamiltonian

To simplify the integration in equation 8, unitless quantities are introduced:

$$\tilde{H} = \frac{H}{H_{\mathrm{sep}}} = \frac{H S^2}{64 \pi^3 d^3} \qquad \tilde{J} = \frac{J}{(2h)^2} = \frac{J S^2}{64 \pi^2 d^2} \tag{10}$$

The three turn phase advance (equation 5) then reads

$$\mu_3 = 6\pi d\left(1 - \sqrt{2\tilde{J}}\cos(3\theta)\right) = 3\pi d\left(\frac{\tilde{H}}{2\tilde{J}} - 1\right) \tag{11}$$

And inserting into equation 8 yields

$$\frac{6\pi d}{\langle\mu_3\rangle} = \frac{1}{2\pi}\oint \frac{1}{1 - \sqrt{2\tilde{J}}\cos(3\theta)}\,\mathrm{d}\theta \tag{12}$$

In the limit of $\tilde{H} \to 0$ the right-hand side of equation 12 equals unity since $\tilde{J} = 0$. Thus, one gets $\langle\mu_3\rangle = 6\pi d$, which is the phase advance per three turns for the linear case. In the limit of $\tilde{H} \to 1$ the integrand has a singularity at the corners of the stable triangle (fixed points) where $\cos(3\theta) = 1$ and $\tilde{J} = 1/2$. The integral becomes infinite and the average three turn phase advance vanishes as expected for a fixed point: $\langle\mu_3\rangle \to 0$.
To carry out the integration over angle for a given value of $\tilde{H}$, one must express $\tilde{J}$ as function of the angle $\theta$ and the invariant $\tilde{H}$. Therefore, we consider the Kobayashi Hamiltonian (equation 3) in unitless quantities

$$\tilde{H} = 6\tilde{J} - 4\sqrt{2}\tilde{J}^{\frac{3}{2}}\cos(3\theta) \tag{13}$$

Solving for $\tilde{J}$ yields 3 solution, of which the following describes the dependence inside the separatrix ($0 < \tilde{H} < 1$):

$$\tilde{J} = \begin{cases} \tilde{H}/6 & \text{if } \cos(3\theta) = 0 \\ \dfrac{6 + R^{1/3}\left(\sqrt{3}i - 1\right) + R^{-1/3}\left(\sqrt{3}i + 1\right)\left(8\tilde{H}\cos^2(3\theta) - 9\right)}{16\cos^2(3\theta)} & \text{otherwise} \end{cases}$$

$$R = 27 - 36\tilde{H}\cos^2(3\theta) + 4\tilde{H}^2\cos^3(3\theta)\left(2\cos(3\theta) - \sqrt{2\cos(6\theta) + 2 - 4/\tilde{H}}\right) \tag{14}$$

The integral in equation 12 can now be computed numerically for a given value of $\tilde{H}$. Figure 2 shows the resulting relation for the average phase advance as function of the invariant $H$. One can see that the detuning effect is very strong close to the separatrix, with the three turn motion freezing in the limit of $H \to H_{\mathrm{sep}}$ as discussed before.

In addition, figure 2 shows the average action $J$ as a function of $H$, which is computed by numerical integration of equation 9. This shows, that the effect of detuning with angle is not negligible and the detuning for a given particle must in fact be considered with respect to $H$. For reference, Figure 3 shows the average phase advance with respect to the average action.

Even though we can not give an analytical result for the integral in equation 12 describing the average three turn phase advance, a Taylor expansion gives a reasonably good approximation for $\tilde{H} < 0.5$ (see figure 4)

$$\langle \mu_3 \rangle = 6\pi d \left( 1 - \frac{2}{3^2}\tilde{H} - \frac{2}{3^3}\tilde{H}^2 + \mathcal{O}\left(\tilde{H}^3\right)\right) \tag{15}$$

To relate the average three turn phase advance $\langle \mu_3 \rangle$ to the average phase advance per turn $\langle \mu \rangle$ we note that

$$\langle \mu_3 \rangle = 3\langle \mu \rangle \bmod 2\pi \tag{16}$$

Recalling the linear case where $\mu = 2\pi q = 2\pi(r + d)$ we can easily find that in general:

$$\langle \mu \rangle = 2\pi r + \langle \mu_3 \rangle / 3 \tag{17}$$

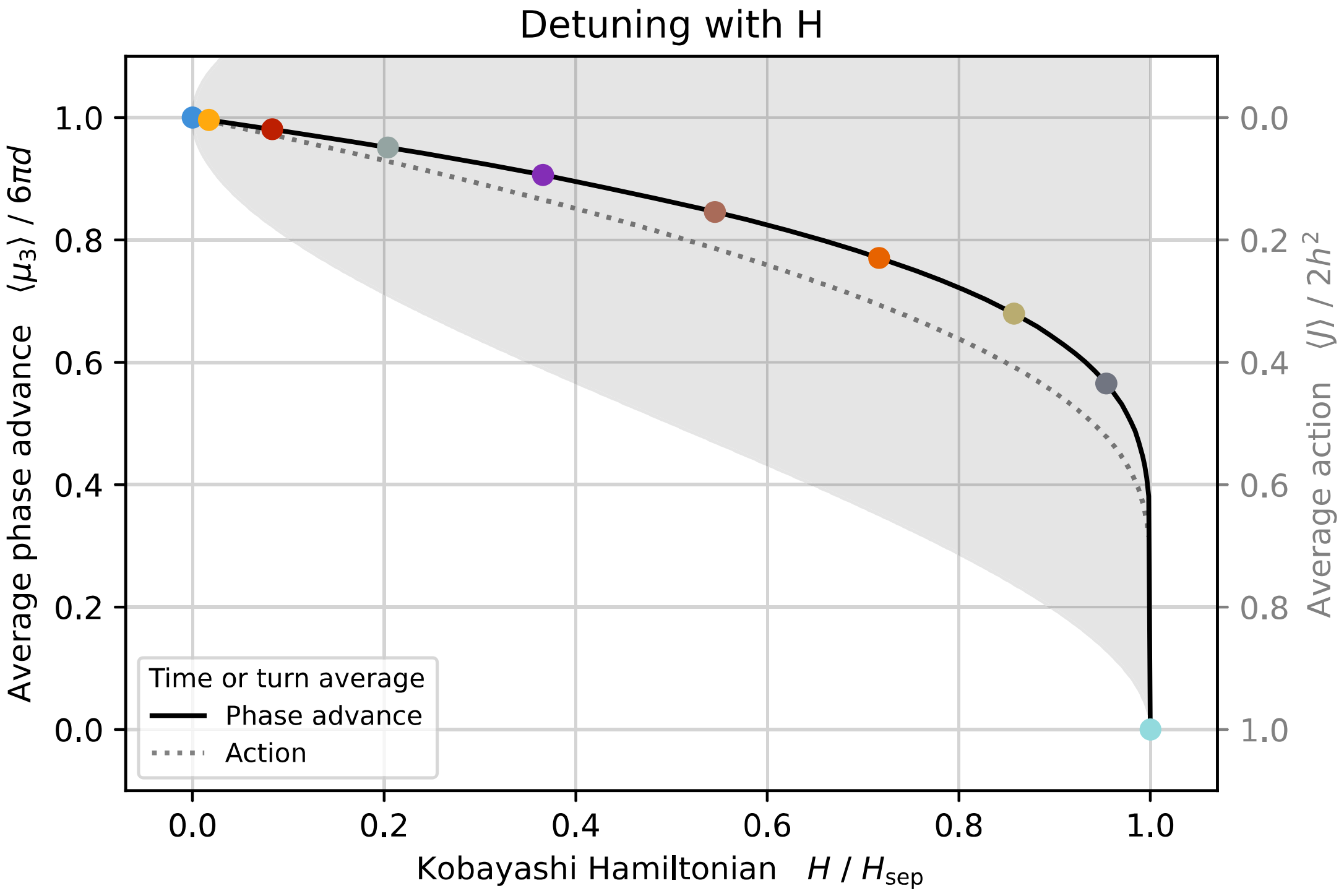

Figure 2: Three turn phase advance $\langle \mu_3 \rangle$ as function of the Hamiltonian $H$. The shaded region shows the range of minimum and maximum three turn phase advance, and the black line is the time average derived in this note. The coloured dots correspond to the distinct values of $H$ presented in figure 1. The dotted line is the average action as function of $H$.

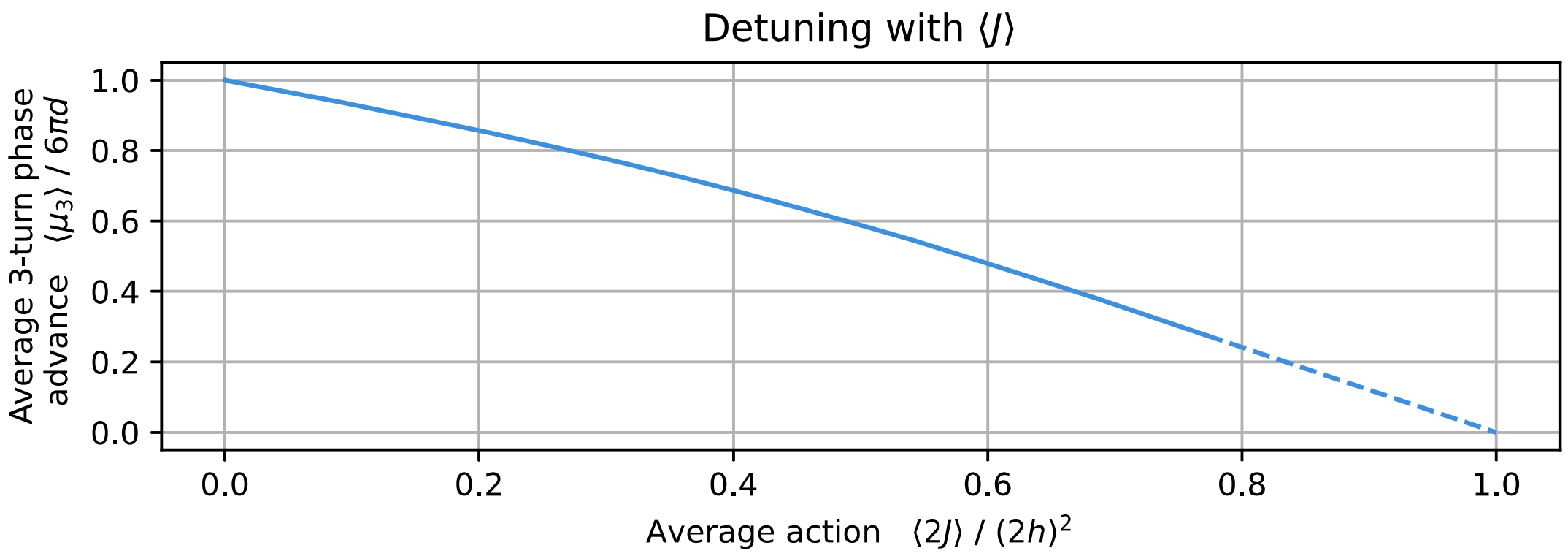

Figure 3: Time average of the three turn phase advance $\langle \mu_3 \rangle$ as function of the time average of the action $\langle J \rangle$. The dashed part is numerically unstable.

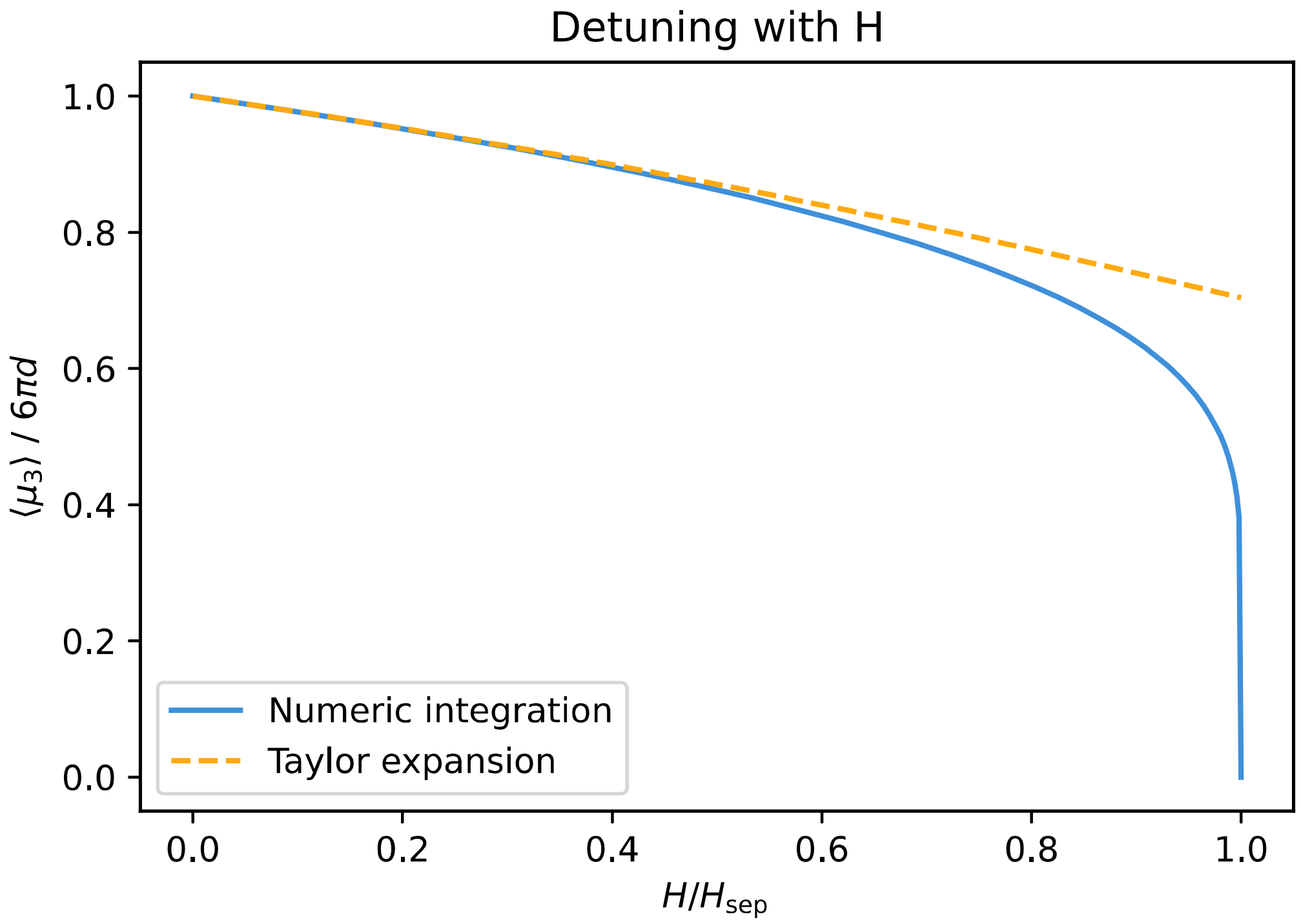

Figure 4: Average phase advance as function of the Hamiltonian $H$ and Taylor approximation.

## 2.3 Comparison to particle tracking

The results found above agree well with results from particle tracking simulations (figure 5, $q = 2/3 + 0.001$, $S = 0.05\,\mathrm{m}^{-1/2}$).

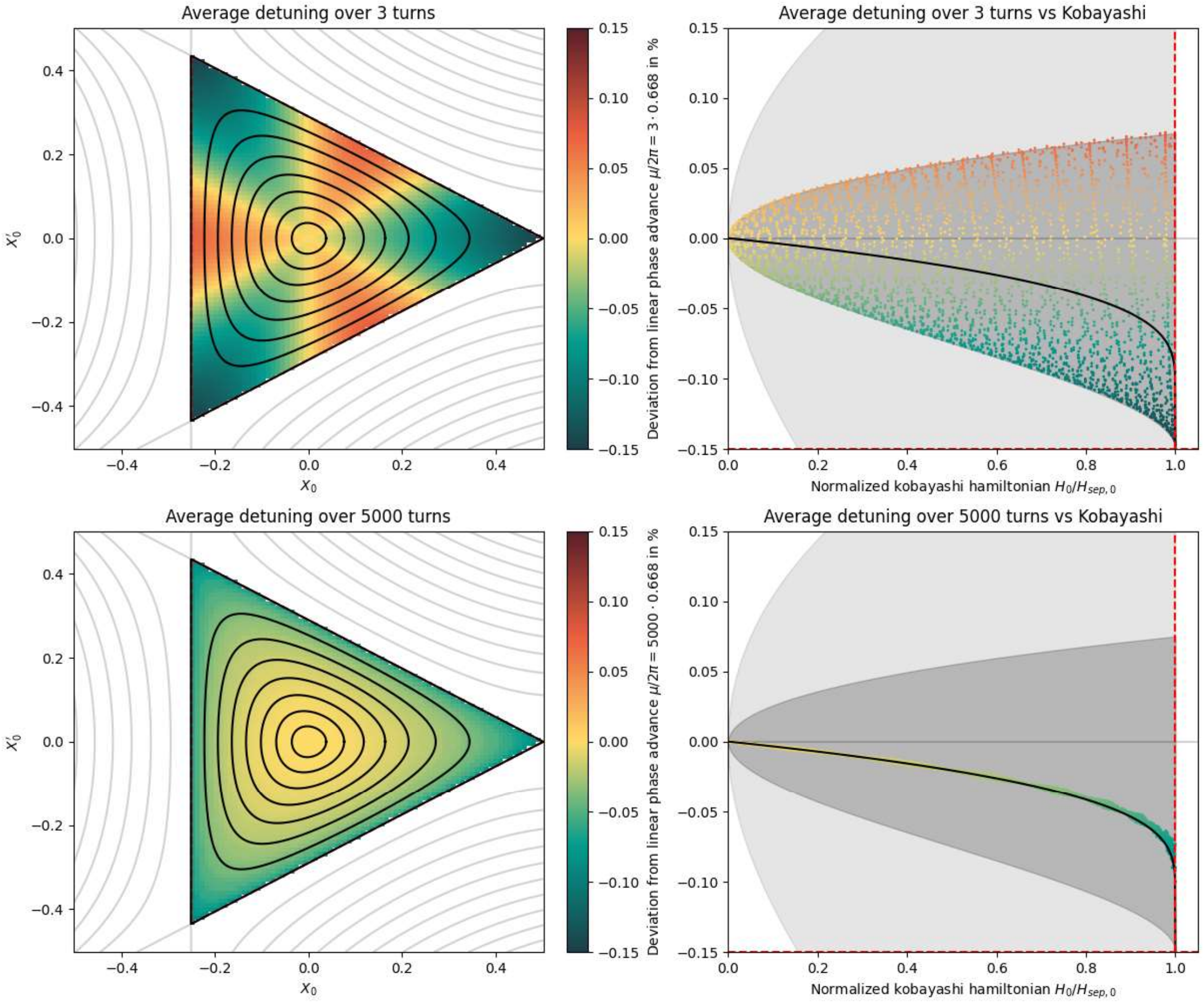

Figure 5: Three turn (top) and average phase advance (bottom) in phase space (left) and as function of $H$ (right). Results from tracking simulations (coloured) are compared to the numerical results as derived above (greyscale).